\documentclass[
    ,final            
  ]
  {aipproc}
\layoutstyle{8x11double}

\usepackage{amssymb}

\newcommand{\be}{\begin{equation}}
\newcommand{\ee}{\end{equation}}
\def\bea{\begin{eqnarray}}
\def\eea{\end{eqnarray}}
\def\ap{\alpha^\prime}
\def\nn{\nonumber}
\def\half{\frac{1}{2}}
\begin{document}

\author{Bin Chen}{
  address={Department of Physics, Peking University, Beijing 100871, P.R. China},
  email={bchen@itp.ac.cn},
}

\author{Miao Li}{
  address={Institute of Theoretical Physics, Chinese Academy of Science, P.O.Box 2735, Beijing 100080, P.R. China},
  email={mli@itp.ac.cn},
  }

\author{Jian-huang She}{
  address={Institute of Theoretical Physics, Chinese Academy of Science, P.O.Box 2735, Beijing 100080, P.R. China},
  email={jhshe@itp.ac.cn},
  }

\title{The Fate of Massive Closed Strings }
\date{2005/7/14}


\begin{abstract}
  We calculate the semi-inclusive decay rate of an average string
state with toroidal compactification in the the superstring
theory. We also apply this calculation to a brane-inflation model
in a warped geometry and find that the decay rate is greatly
suppressed if the final strings are both massive and enhanced for
massless radiation.
\end{abstract}

\maketitle


\bigskip

\section{Introduction}

Our study of the massive closed string states was motivated by the
renewed interests in cosmic string\cite{cosmic, cosmica, cosmicb},
and also by a desire to understand the mysterious open/closed
duality in tachyon condensation\cite{sen, openclose}. The revival
of the cosmic string resides in the fact that in the low-energy
models the cosmic strings could be created after inflation. In a
class of string-inspired models on inflation, the tachyon
condensation plays a fundamental rule. It has been shown that some
strings, including fundamental string and solitonic strings, could
be created after tachyon condensation. They are sufficiently
stable and a few of them may survive and be observable in near
future through gravitational wave detectors such as LIGO and LISA.
Furthermore, in the decay products of the tachyon condensation,
there are very massive closed string states, which could be an
origin of the cosmic string. However these massive fundamental
string states are not stable and may decay to other massive or
massless states. In order to understand the fate of a typical
massive F-string state better, it is essential to study its decay
systematically. On the other hand, even without considering its
phenomenological implication, the study of the decay of a typical
massive string state is an interesting issue on its own right in
string theory. To understand the tachyon condensation and the
mysterious open/close string duality, the study of the fate of
these massive string states is necessary.

The decay of massive strings is an old problem of string theory
and it is generally difficult to extract detailed knowledge from
conventional calculations mainly due to the exponentially growing
state density. States at the same mass level may be very
different, especially their decay properties. Fortunately in a
realistic situation, such as the rolling tachyon case, details of
string states are not important, and  we are only interested in
some sort of averaged decay rates. In an algebraic approach, J.L
Manes considered in \cite{ma} the emission spectrum of the strings
of all mass levels for bosonic (open and closed) strings. It was
found there that the decay rates are universal, in the sense that
the leading terms depend only on the mass level of the strings
involved, and the more detailed properties of the states affect
only the sub-leading terms.

To study the stringy properties relevant for cosmology, we need to
consider the situation
 where some of the space-time dimensions are compactified. And in the realistic
 models, the superstrings rather than the bosonic strings are used.
Therefore we will try to study the closed string decay in
superstring in the flat spacetime with compactifications and then
in a realistic model\cite{kklmmt} with warped compactification.
The work reported here is based on the paper\cite{CLS}.

\section{Closed String Decay: with Compactification}

We firstly study the emission spectrum of massive closed bosonic
strings in the case where
 $d_{c}$ out of the total $D$ dimensions are compactified on a flat torus.
A closed string state is characterized by  its KK momentum numbers
$n_{i}$, winding numbers $w_{i}$, and its mass $M$ and momentum
$k$ in the noncompact dimensions. The mass shell condition reads {
\be \ap M^2 = 4 (N_{R}-1)+ \ap Q_{+}^2 = 4 (N_{L}-1)+ \ap Q_{-}^2
\nn\ee} where $Q_{\pm} = \sum {n_{i}\over R_i}\pm {w_{i}R_i\over
\ap}$ , and $i=1,...,d_{c}$.

 Consider the  process
 \be
 (M,n_i,w_i)\to (m,n_{1i},w_{1i})+(M_2,n_{2i},w_{2i})
 \nn\ee
   In the following, we set $\ap = \half$.

We are interested in the averaged semi-inclusive two-body decay
rate: for the initial string and one of the two final strings, we
average over all states of some given mass, winding and KK
momentum. Only the other string's state is fully specified (by
keeping explicit its vertex operator). This decay rate can be
written as
 \be
 \Gamma_{semi-incl} = {A_{D-d_{c}}\over
{M^2}}g_{c}^2 {F_{L}\over{{\cal G}(N_L)}} {F_{R}\over{{\cal
G}(N_R)}}k^{D-3-d_c}\prod_i^{d_{c}} R_{i}^{-1}
\ee
 with closed
 string coupling $g_c$, compactification radius $R_i$, and $F_L$
and $F_R$ are given by
 \bea
 F_{L}&=& \sum_{\Phi_i}\sum_{\Phi_f} \mid\langle
 \Phi_f \mid V_{L}(n_{1i},w_{1i},k)\mid\Phi_i \rangle \mid^2 \nn \\
F_{R}&=& \sum_{\Phi_i}\sum_{\Phi_f}\mid
 \langle \Phi_f \mid  V_{R}(n_{1i},w_{1i},k)\mid\Phi_i \rangle \mid^2
 ,
 \nn\eea
which can be written as  \be F_L = \oint _C {dz\over z}z^{-N_{L}}
\oint _{C'} {dz' \over z'}z'^{-N_{2L}} {\rm Tr} [z^{\hat
N}V_{L}^\dagger(k,1) z'^{\hat N}V_{L}(k,1)] .\nn\ee In the
following, we define ${ w=zz'}$, and ${ v=z'}$.

Note that  \bea & & \int d^{D}p {\rm Tr} [z^{ L_{0} } V^\dagger
(k,1){z'}^{ L_{0} } V(k,1)] \nn\\
&=&f(w)^{2-D}({{-2\pi}\over{\ln w}})^{D/2}< V^\dagger (k,1) \
V(k,v) >_c, \nn\eea where  \be
f(w)=\prod_{n=1}^{\infty}(1-w^n)^{2-D}, \nn\ee and $<\cdots>_c$
denote the correlators on the cylinder.  Thus the $master formula$
for the amplitude squared is  \bea F_L &=&\oint _{C_w}{dw\over
w}w^{-N_{L}}f(w)^{2-D} {{\cal
I}_{N_{L}-N_{2L}}(w)}\\
 {\cal I}_n &=& \oint _{C_v}{dv\over
v}v^n <{V'}_{L}^\dagger(k_L,1)V'_L(k_L,v)>_c \nn\eea
 where the contours
satisfy $|w| < |v| < 1$, for $|z|,|z'| < 1 $.

The key point here is that there exist a recursion
relation\cite{ma}, \be {\cal I}_{n+m_L^2}(w) = w^{n+m_L^2}[{\cal
I}_n (w)+n+{m_L^2\over 2}],\nn\ee which allow us to write the
${\cal I}_n$ into a series summation. Straightforwardly, the
master formula could be written as  \be F=\sum_{p=1}^A
(n-m_L^2(p-\half)){\cal G}[N_L-np+\half
m_L^2(p^2-p)]+F_{NU},\nn\ee
 where $n=N_L-N_{2L}$, and
 \be F_{NU}=\oint _{C_w}{dw\over w}w^{-N}f(w)^{2-D}{\cal I}_\nu
(w)w^{\nu A+\half m_L^2(A^2+A)}\nn\ee and the generating function
for the mass level density  \be {\rm Tr} w^{\hat
N}=f(w)^{2-D}=\sum_{N=0}^{\infty}{\cal G}(N)w^N .\nn\ee
 has been
used.

In $F$, all terms except the last one are universal, in the sense
that they does not depend on the details of any of the three
string states involved. The last term $F_{NU}$ is non-universal.
The contribution from $F_{NU}$ to $F$ is generically negligible,
in comparison with the universal part contribution, except the
case $A=0$ which happens when the emitted states carry a large
fraction of the total mass.

One important observation is that we have inequality \be
\sqrt{N_{1L}}+\sqrt{N_{2L}} \leq \sqrt{N_{L}}.\nn\ee The equality
saturate when  \be k=0,\ \ \ \ \ \ {M \over Q_-}={M_2\over
Q_{2-}},\nn\ee where $k$ is the momentum in the noncompact
directions and $M_2, Q_{2-}$ are the quanta of the outgoing string
states with fixed level $N_{2L}$. The same inequality holds in the
right-mover.  Thus if generically $N_L
> N_{2L}>{N_L\over 3}$, the first term dominates the whole summation in $F_L$, and the
other terms will be neglected to get  \be F_L \approx
(N_L-N_{2L}-\half m_L^2){\cal G}(N_{2L}).\nn\ee $F_R$ can be
carried out in the same way.

 We can get the total decay rate for
decays into arbitrary states of given mass, winding and KK
momentum by simply multiplying  by the state density ${\cal
G}(N_{1})$ \bea & & \Gamma
[(M,n_i,w_i)\to(m,n_{1i},w_{1i})+(M_2,n_{2i},w_{2i})] \nn\\
&\approx& A_{D-d_c}{g_c^2\over M^2}{\cal N}_L {\cal N}_R {\cal
G}_L {\cal G}_R k^{D-3-d_c}\prod _i ^{d_{c}} R_{i}^{-1},\nonumber
\nn\eea where  \be {\cal N}_L=N_L-N_{2L}-\half m_L^2\ ,\ \ \ {\cal
N}_R=N_R-N_{2R}-\half m_R^2,\nn\ee and  \be {\cal G}_L={{{\cal
G}(N_{1L}){\cal G}(N_{2L})}\over{\cal G}(N_L)}\ ,\ \ \ {\cal
G}_R={{{\cal G} (N_{1R}){\cal G}(N_{2R})}\over{\cal
G}(N_R)}.\nn\ee

As long as $N_{1L}\gg 1$ and $N_{2L}\gg 1$, we can write  \be
{\cal G}_L \sim (2\pi T_H)^{-{{D-1}\over 2}}({N_{1L}N_{2L}\over
{N_L}})^{-{{D+1}\over 4}}e^{-\sqrt{2}t_L/T_H},\nn\ee with the
Hagedorn temperature $T_H={1\over\pi}\sqrt{3\over{D-2}}$ and  \be
t_L=\sqrt{N_{L}}-\sqrt{N_{1L}}-\sqrt{N_{2L}}. \nn\ee We have in
the above restored the multiplicative constant in front of the
state density.

From the  inequality above, we know that the dominant decay
channel is the original incoming string break into two strings
with the same  \be \alpha={M\over Q_-}={m\over Q_{1-}}={M_2\over
Q_{2-}}\nn\ee and without releasing any kinetic energy $k=0$.


The total decay rate is
  \bea
 & &\Gamma_m(M,n_i,w_i)\nn\\
 &=&\sum_{N_2}\sum_{(n_{1i},n_{2i})}\sum_{(w_{1i},w_{2i})}\Gamma
[(M,n_i,w_i)\to (m,n_{1i},w_{1i})+(M_2,n_{2i},w_{2i})].\nn\eea The
decay rate can be well approximately factorized into three parts
\be \Gamma_m(M,n_i,w_i)=\Gamma^k \sum_{(n_{1i},n_{2i})}{\cal
G}^{KK}\sum_{(w_{1i},w_{2i})}{\cal G}^w,\nn\ee where we make
saddle point approximation for the noncompact momentum, that is we
let $k=0$, or $M_2=M-m$,
 when considering the KK modes and winding modes.

  In general, the
contribution from the KK-modes and windings could be just a
constant factor and the dependence on the momentum $k$ could be
encoded in  \be ({mM_2 \over 4})^2\exp\big({-({\alpha \over
\sqrt{\alpha^2-1}}+{\beta \over \sqrt{\beta^2-1}}){k^2\over
2T_H}{M\over 2mM_2}}\big),\nn\ee where $\beta={M \over Q_+}$. Note
that if we measure the energy of the one of the decay string with
mass $m$, we find a temperature:  \be T_m={T_H\over 2}({\alpha
\over \sqrt{\alpha^2-1}}+{\beta \over \sqrt{\beta^2-1}})^{-1}(1-{m
\over M}).\nn\ee This reflects both the recoil effect and the
influence of the KK-modes and windings.

It is remarkable that when $\alpha=1$ or $\beta=1$, the decay is
completely suppressed. This interesting limit corresponds to the
BPS condition $M=Q_-$ or $M=Q_+$ in the superstring case. As we
show below, all the discussion here could be applied to the
superstring. Therefore, the fact that the BPS string states are
stable is reflect in the above special limit. Actually, the
suppression not only happens in the decay to the final massive
strings, it also happens in the thermal radiation of the massless
particles. Moreover, from above, the decay of the near BPS string
states are greatly suppressed.

Taking into account of the measure of integral, we have  \be
\Gamma^k= A'_{d_c} g_c^2 {(2\pi T_H)}^{-D+1}{M\over m}
m_R^{-{{D-1+d_c}\over2}}\int_0^{\infty}
t^{{D-3-d_c}\over2}e^{-{t\over T_H}}dt, \nn\ee where  \be
t=({\alpha \over \sqrt{\alpha^2-1}}+{\beta \over
\sqrt{\beta^2-1}}){k^2\over 2T_H}{M\over 2mM_2}\nn\ee and
$m_R={{m(M-m)}\over M}$, and $A'_{d_c}\approx
{\mbox{const.}}\prod_{i=1}^{d_c} R_i^{-1}$. Thus including the
phase space factor, we have a Maxwell-Boltzmann distribution for
$t$, which could be taken as a measure of the total kinetic
energy.

The mean kinetic energy released per decay could be characterized
by \be <t>={D-1-d_c \over 2}T_H,\nn\ee which is independent of the
masses and satisfies the equipartition principle in $D-1-d_c$
noncompact spatial dimensions.

 Carrying out the integral, which is just a $\Gamma$-function, we get
\be \Gamma ^k\approx A'_D g_c^2{M\over
m}m_R^{-{{D-1+d_c}\over2}},\nn\ee with $A'_D= const.
A'_{d_c}{T_H}^{-{(D-1+d_c)\over2}}$.

The emission rate of strings of mass $m$ is
 \be
{d\Gamma(m)\over{dm}}={{d\Gamma^k(m)}\over{dm}}
\sum_{(n_{1i},n_{2i})}{\cal G}^{KK}\sum_{(w_{1i},w_{2i})}{\cal
G}^w,\nn\ee where  \be {{d\Gamma^k(m)}\over{dm}}=\Gamma^k
\rho(m)=A''_D g_c^2Mm_R^{-{{D-1+d_c}\over2}},\nn\ee with
$A''_D={1\over4}A'_D$. It can easily be checked that if we set
$d_c=0$,
 the above result reduces to the flat space case.

 Next let us consider the decay rate of a typical massive type II string
state in the 10-d dimensional Minkowski space. The basic quantity
 to compute is one piece of the semi-inclusive amplitudes squared,
 \be F={1\over {\cal G} (N)}\sum_{\Phi_i}\sum_{ \Phi_f} \mid
 \langle \Phi_f \mid  V(k)\mid\Phi_i \rangle \mid^2. \nn\ee

 One observation is that NS emission is the dominant decay channel
for $N$ large, and R sector is suppressed by $1\over N$, with $N$
being the initial string mass level. For the NS channel, the key
point is to
 check whether there is still a recursion relation when the fermions are
 included. The above tree level trace could still be converted to
 a correlator on the cylinder. What is new for the superstring case is that due to the presence
of fermions, the correlator  is a summation over spin structures.
The final result is that we have the same recursion relation for
${\cal I}_n$  as in the bosonic string and the only change is to
replace space-time dimension $26$ by $10$. Consequently, the study
on the decay of the massive string states in the bosonic string
theory with toroidal compactification could be carried over to the
superstring case directly.

\section{A realistic model}

The results on the toroidal compactifications can be used to
estimate the massive string decay rates in the more realistic
 KKLMMT model, where the type IIB strings live in a highly warped throat with Klebanov-Strassler
geometry.

A general warped geometry  \be
ds^2=H^{-\half}(Y)\eta_{\mu\nu}dX^\mu dX^\nu +H^\half
(Y)g_{ij}dY^i dY^j\nn\ee will induce a potential  \be
V(Y)={1\over{2\pi \ap H^{1\over2}(Y)}}\nn\ee on the string
worldsheet, this potential confines the string to a small region
in the compact dimension near the position where this potential
has a minimum. Fluctuations in these compact dimensions  \be
<Y^iY^i>={\ap\over2}\ln[1+{1\over{2\pi {\ap}^2 H^\half
(0)\partial_i^2 V(0)}}], \nn\ee provide an effective
compactification volume for strings living in. When $\partial_i^2
V$ is small in string units, that is, the geometry varies slowly
on the string scale, we can look upon the strings in such a warped
geometry as in a box with the effective compactification volume,
and the flat space calculations above applies.

For the warped geometry, what we have discussed represents results
seen by a 10-dimensional observer. And the quantities measured by
the 4-dimensional observer, in our case the mass and the momentum,
get red-shifted. Thus we should multiply $M, m,M_2$ and $k$ all by
a red-shift factor $H^{1\over4}(0)$. Now the mass shell conditions
read like  \be \ap M^2H^{\half}(0)=4(N-1).\nn\ee

 The emission rate of a string with mass $m$
from a typical string of mass $M$ in the Klebanov-Strassler
geometry can finally be estimated to be  \be
{{d\Gamma_{KS}(m)}\over{dm}}\approx A_{KS}g_c^2
H^{-{13\over8}}(0)Mm_R^{-{15\over2}},\nn\ee with $A_{KS}=
{1\over{16\pi^5}}{{V_{min}}\over{V_{\perp}}}$.

With $H^{-{1\over4}}(0)\sim 10^{-4}$, we see that the emission
rate is suppressed by a factor of order $10^{-26}$ in this warped
geometry!  Thus in this warped geometry, the decay channels for a
massive string to two massive strings are greatly suppressed.

 We note that the situation is different if
one of the two final strings is massless. It is found  that the
decay rate is proportional to the mass of the initial string. So
including the redshift factor, it can be written as  \be
\Gamma_{massless}\sim g_c^2 M H^{1\over 4}(0),\nn\ee
 from which we see that this decay
channel is enhanced by order $10^4$. Thus in such warped
geometries, massive strings may lose their energy mainly through
gravitational radiation.

\section{Conclusion}

 We calculate the cross section of the decay of a
massive closed string state into two massive closed string states;
The decay shows some novel features: KK modes and windings
proportional to the masses preferred;  the superstring share the
same recursion relation with the bosonic string; In a realistic
model with warped geometry, the massless decay is enhanced by the
redshift factor.

\vspace{1cm}

{\bf Acknowledgments}

This research project was supported by a grant from CNSF and a
grant from CAS. We would like to thank J.L Manes for valuable
comments. BC would like to thank the invitation from PASCOS-05 to
present this work.

\end{document}